\documentclass[twocolumn]{article}

\usepackage[english]{babel}

\usepackage[letterpaper,top=2cm,bottom=2cm,left=2cm,right=2cm,marginparwidth=1.75cm]{geometry}

\usepackage{amsmath}
\usepackage{amssymb}
\usepackage{physics}
\usepackage{graphicx}
\usepackage{outlines}
\usepackage[colorlinks=true, allcolors=blue]{hyperref}
\usepackage{cleveref}
\usepackage{enumitem}
\usepackage[normalem]{ulem} 

\usepackage{float}
\usepackage{caption, setspace}
\usepackage{subcaption}
\usepackage{placeins}
\usepackage[backend=biber,style=nature]{biblatex}
\title{Effective Data-Driven Collective Variables for Free Energy Calculations from Metadynamics of Paths}
\author{
Lukas Müllender\footnote{Department of Applied Physics, Science for Life Laboratory, KTH Royal Institute of Technology, Stockholm, Sweden} \footnote{Computational Biomedicine, Institute of Advanced Simulations IAS-5/Institute for Neuroscience and Medicine INM-9, Forschungszentrum Jülich GmbH, Jülich, Germany} \footnote{Department of Physics, RWTH Aachen University, Aachen, Germany}
\and
Andrea Rizzi\footnotemark[2] \footnote{Atomistic Simulations, Italian Institute of Technology, Genova, Italy}
\and
Michele Parrinello\footnotemark[4]
\and
Paolo Carloni\footnotemark[2] \footnotemark[3] \footnote{Universitätsklinikum, RWTH Aachen University, Aachen, Germany} \textsuperscript{,1}
\and
Davide Mandelli\footnotemark[2]  \textsuperscript{,2}
}

\begin{document}
\maketitle
\footnotetext[1]{Email: p.carloni@fz-juelich.de}
\footnotetext[2]{Email: d.mandelli@fz-juelich.de}

\textbf{ABSTRACT. A variety of enhanced sampling methods predict multidimensional free energy landscapes associated with biological and other molecular processes as a function of a few selected collective variables (CVs). The accuracy of these methods is crucially dependent on the ability of the chosen CVs to capture the relevant slow degrees of freedom of the system. For complex processes, finding such CVs is the real challenge. Machine learning (ML) CVs offer, in principle, a solution to handle this problem. However, these methods rely on the availability of high-quality datasets -- ideally incorporating information about physical pathways and transition states -- which are difficult to access, therefore greatly limiting their domain of application. Here, we demonstrate how these datasets can be generated by means of enhanced sampling simulations in trajectory space via the metadynamics of paths~(Mandelli et al., 2020) algorithm. The approach is expected to provide a general and efficient way to generate efficient ML-based CVs for the fast prediction of free energy landscapes in enhanced sampling simulations. We demonstrate our approach with two numerical examples, a two-dimensional model potential and the isomerization of alanine dipeptide, using deep targeted discriminant analysis as our ML-based CV of choice.}
\newpage

Enhanced sampling (ES) methods \cite{heninEnhancedSamplingMethods2022a,valssonEnhancingImportantFluctuations2016} are a powerful tool to investigate rare events in molecular systems, such as conformational changes of large biomolecular complexes, drug binding to receptor targets or phase transitions in materials \cite{delemotteFreeenergyLandscapeIonchannel2015,tiwaryKineticsProteinLigand2015,blaakCrystalNucleationColloidal2004}.
To obtain the free energy landscape describing these complex phenomena, 
a large class of ES methods work under the assumption that a few collective variables (CVs), functions $s(\mathbf{R})$ of the atomic coordinates, exist that are able to provide a concise description of the transformation of interest. 
An external potential $V(s)$ can then be defined, able to drive the rare transitions and allowing a reconstruction of the free energy profile. These methods include, among many others, umbrella sampling \cite{torrieNonphysicalSamplingDistributions1977}, hyperdynamics~\cite{voterHyperdynamicsAcceleratedMolecular1997}, well-tempered or OPES-based metadynamics~\cite{laioEscapingFreeenergyMinima2002,barducciWellTemperedMetadynamicsSmoothly2008b,invernizziRethinkingMetadynamicsBias2020a}, adaptive biasing force~\cite{darveAdaptiveBiasingForce2008}, or variationally enhanced sampling~\cite{valssonVariationalApproachEnhanced2014b}. We note here, that in cases where sufficiently long unbiased trajectories are available, researchers have designed general and elegant methods to describe the thermodynamics of a system without making use of CVs~\cite{lindorff-larsenHowFastFoldingProteins2011,sormaniExplicitCharacterizationFreeEnergy2020b}. However, being able to access the Boltzmann distribution from unbiased MD simulations is the exception rather than the rule.

The success of CV-based ES algorithms relies on the highly nontrivial choice of the CVs, which must be able not only to discriminate between the different metastable states, but also, and most importantly, to describe the progress of the reaction. Recently, machine learning-based (ML) methods have been shown to be effective in delivering CVs that fulfill these two criteria~\cite{chenCollectiveVariablebasedEnhanced2021,mendelsCollectiveVariablesLocal2018,bonatiDataDrivenCollectiveVariables2020,perez-hernandezIdentificationSlowMolecular2013a,bonatiDeepLearningSlow2021,chenMolecularEnhancedSampling2018,sultanAutomatedDesignCollective2018,ribeiroReweightedAutoencodedVariational2018,sunMultitaskMachineLearning2022,hooftDiscoveringCollectiveVariables2021}. 
However, as in any machine learning approach, the results depend dramatically on the quality of the underlying data. This leads to a chicken-and-egg problem: for good results, one ideally needs data on the relevant metastable states and transitions between them, which, in turn, would require knowledge of a CV that allows their thorough sampling~\cite{rohrdanzDiscoveringMountainPasses2013}. As a result, most data-driven CV approaches still struggle to adequately accelerate the important motions in complex systems. To improve their efficiency, it has been previously recognized that including data from the transition state plays an important role in promoting the adequate description of the transition dynamics~\cite{rayDeepLearningCollective2023}. 

\begin{figure*}[!htb]
    \centering
    \includegraphics[width=\textwidth]{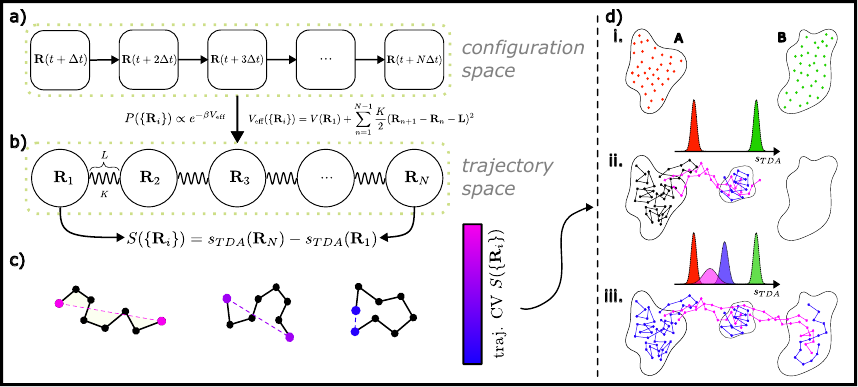}
    \caption{(a)  A discrete stochastic trajectory consists of the time series $\{{\bf R}_n\}$ of configurations visited sequentially by the system. (b) To each discretized trajectory can be assigned a well-defined Boltzmann-like statistical weight $e^{-\beta V_{\rm eff}(\{{\bf R}_n\})}$. The effective potential is isomorphic to the potential energy of an elastic polymer. (c) One can generate all possible discretized trajectories of a physical system by evolving the dynamics of a fictitious polymer subject to the force $\mathcal{F}_n=-\nabla_{{\bf R}_n}V_{\rm eff}$. Metadynamics can be used to accelerate the sampling of reactive pathways by choosing an appropriate collective variable defined in trajectory space. Here, we adopt as CV$_t$ the generalized polymer end-to-end distance $S$ of equation~\ref{eq:traj_cv}, defined as the difference between the Deep-TDA CV$_c$ evaluated in the final and in the starting configuration of the polymer. (d) Iterative procedure for DeepTDA CV$_c$ training. (i) Unbiased training data sets are generated in the initial (A) and final (B) states. A DeepTDA CV$_c$ $s_{TDA}$ is trained to map the two datasets into well-separated Gaussian distributions. (ii) A MoP run is performed and the generated paths are classified as trapped (blue and black) and reactive (violet) paths, based on the value of $|S|$. New datasets are identified from newly discovered metastable states and from configurations occupying the transition state region. A multi-state DeepTDA CV$_c$ is trained to map these configurations into well-separated distributions and a new end-to-end distance CV$_t$ is built. (iii) The desired outcome of a MoP run using the refined end-to-end CV$_t$ is depicted, showing sampling of trapped paths in all metastable states (in blue) as well as reactive pathways connecting different basins (in violet). This data can be used to design an optimal CV$_c$ in configuration space to converge the free energy landscape in a standard metadynamics simulation.}
    \label{fig:mop_iteration}
\end{figure*}

To harvest this essential data from the transition path ensemble, we shift our attention from enhanced sampling methods in configuration space to approaches focused on the direct sampling of the transition pathways. Among many such methods based on the statistical mechanics of trajectories~\cite{onsagerFluctuationsIrreversibleProcesses1953,prattStatisticalMethodIdentifying1986,mandelliModifiedNudgedElastic2021}, we consider the recently developed metadynamics of paths (MoP) algorithm~\cite{mandelliMetadynamicsPaths2020a}. In contrast to the popular transition path sampling~\cite{dellagoTransitionPathSampling1998,bolhuisTransitionPathSampling2002}, which has also been used for the identification of  CVs~\cite{maAutomaticMethodIdentifying2005,bestReactionCoordinatesRates2005,sunMultitaskMachineLearning2022,hooftDiscoveringCollectiveVariables2021}, MoP allows for the unconstrained exploration of multiple reactive paths connecting metastable states without the need for an initial path guess. This is achieved by performing metadynamics simulations in the space of all trajectories, making use of special collective variables defined in \textit{trajectory space} (CV$_t$ hereafter). As we will see below, a crucial property of MoP is its robustness in sampling the transition path ensemble with respect to a suboptimal choice of this CV$_t$, considerably mitigating the chicken-and-egg problem described above.

We show that the data obtained from MoP can be used to train an efficient ML-based CV in configuration space (CV$_c$ hereafter) to speed up standard metadynamics simulations. Here, among many different, powerful ML approaches, we use the deep targeted discriminant analysis (DeepTDA) supervised learning approach\footnote{Note that the choice of DeepTDA was made here mainly because of its simplicity and straightforward interpretation. However, the data generated by MoP could be used to train many other data-driven methods~\cite{chenMolecularEnhancedSampling2018,perez-hernandezIdentificationSlowMolecular2013a,bonatiDeepLearningSlow2021,sultanAutomatedDesignCollective2018,hooftDiscoveringCollectiveVariables2021,sunMultitaskMachineLearning2022}.}~\cite{trizioEnhancedSamplingReaction2021} previously developed to build CV$_c$s. Briefly, DeepTDA trains a classifier that discriminates configurations belonging to different metastable states by mapping them into well-separated, user-defined locations in latent space. This approach can be used to incorporate not only data from multiple metastable states, but also from reactive trajectories connecting them~\cite{trizioEnhancedSamplingReaction2021,rayDeepLearningCollective2023}. The mapping is done such that the resulting one-dimensional CV$_c$ describes the system's progress from one basin to the other through the transition state region (see figure~\ref{fig:mop_iteration}b, and Materials and Methods for more details).

In practice, our proposed, iterative protocol (figure~\ref{fig:mop_iteration}b) consists of the following steps:
\begin{enumerate}[label=\textbf{Step \arabic{*}:}, align=left]
    \item Standard MD simulations lead to kinetically trapped conformations in the (assumed to be known) initial and final basins (figure~\ref{fig:mop_iteration}b). A CV$_c$ is obtained by training a DeepTDA model to discriminate these 2 states
    \item Starting from such CV$_c$, a CV$_t$ is built and MoP simulations are performed
    \item The resulting trajectories are analyzed to identify newly discovered metastable states and reactive paths, and a new DeepTDA CV is trained including these data. If the latest MoP simulation found a path between initial and final states, the algorithm ends: it provides a complete map of the intermediate states and pathways of the molecular transform, which generally allows building efficient CV$_c$. Otherwise, step 2 is repeated with the new CV$_c$.
\end{enumerate}
The method is designed to iteratively refine both CV$_c$ and CV$_t$. More details on how the latter are constructed are provided below.

The paper is organized as follows: after an introduction to MoP and the definition of CV$_t$, we apply our iterative protocol (i) to a 2D model potential, used to test its applicability to multi-state systems, and (ii) to the isomerization of alanine dipeptide in vacuum, which, despite its simpler two-state nature, provides a non-trivial test case on a molecular system.

\paragraph{Metadynamics of paths}
In standard MD simulations, a discrete trajectory -- consisting of the time series $\{{\bf R}_n\}_{n=1,N}$ of configurations visited by the system -- is generated in a {\it sequential} manner, due to the inherent seriality of the time evolution process (see Fig.~\ref{fig:mop_iteration}a). MoP circumvents this problem -- which rests at the base of the poor scaling of MD algorithms -- and achieves parallelization in  time by sampling directly from the phase space of all possible trajectories. The method applies to stochastic (Brownian) trajectories and exploits the isomorphism between the path probability distribution, $p[\mathcal{A}({\bf R}(t))]$, and the Boltzmann distribution of a fictitious elastic polymer (see Fig.~\ref{fig:mop_iteration}b):
\begin{gather}
    p[\mathcal{A}({\bf R}(t))] = \exp\left[-\beta V_{\rm eff}(\{{\bf R}_n\})\right] \\
    V_{\rm eff} = U({\bf R}_1)+\sum_{n=1}^{N-1}\frac{K}{2}\left({\bf R}_{n+1}-{\bf R}_{n}-{\bf L}_{n}\right)^2
    \label{eq:mop}
\end{gather}
In this equation, $\mathcal{A}$ is the Onsager-Machlup action~\cite{onsagerFluctuationsIrreversibleProcesses1953}, which is a functional of the (discretized) Brownian trajectory ${\bf R}(t)={\bf R}_1\rightarrow{\bf R}_2\rightarrow\dots\rightarrow{\bf R}_N$. $\beta=1/k_BT$, while $K=m\nu/2\Delta t$ and ${\bf L}_n=(\Delta t/m\nu){\bf F}_n$ are the effective spring constant and equilibrium length that depend on the physical parameters of the underlying Brownian dynamics: temperature $T$, mass $m$, damping coefficient $\nu$, time step $\Delta t$ (see Materials and Methods for details). $U$ is the potential energy of the system and ${\bf F}_n=-\nabla U({\bf R}_n)$ is the physical force acting on the $n$th configuration.

Finite temperature MD simulations of the polymer are performed by computing the fictitious forces ${\bf\mathcal{F}}_n=-\nabla_{{\bf R}_n} V_{\rm eff}$ acting on each configuration and are used to generate discretized trajectories distributed according to $p[\mathcal{A}({\bf R}(t))]$. 
Metadynamics, in turn, can be used to focus the sampling on the important reactive trajectories connecting metastable states. This requires defining CV$_t$ in trajectory space.

Following Ref.~\cite{mandelliMetadynamicsPaths2020a}, we define our CV$_t$ as the generalized end-to-end distance
\begin{equation}
    S(\{\mathbf R_n\})=s(\mathbf R_N)-s(\mathbf R_1),
    \label{eq:traj_cv}
\end{equation}
where, in this work, $s(\mathbf R)$ is a DeepTDA CV$_c$. The rationale for this specific choice of $S$ is that it allows discriminating between elongated polymers (large values of $|S|$), which are likely to represent \textit{reactive} trajectories, from kinetically \textit{trapped} ones (with a low value of $|S|$), thus aiding in the discovery of new metastable states.

We note that reactive trajectories obtained from this method tend to spend more time in proximity to the transition states. This happens because the equilibrium spring constants are proportional to the physical force vector (${\bf L}_n\propto{\bf F}_n$) and, therefore, tend to zero close to the stationary points of the potential energy surface, including the unstable saddle points~\cite{mandelliModifiedNudgedElastic2021}. This feature increases the amount of data generated on transition states that can be used to train an efficient ML CV$_c$.

\paragraph{Two-Dimensional Model Potential}
\begin{figure}[!ht]
    \centering
    \includegraphics[width=\linewidth]{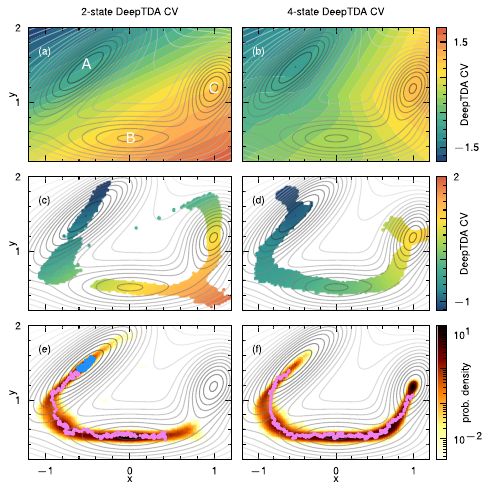}
    \caption{(a) Colored map showing the values of the initial 2-state DeepTDA CV$_c$. Initial, intermediate and final states are labeled as A, B and C, respectively. (b) Colored map of the 4-states DeepTDA CV$_c$. Scatter plots of $x,y$ coordinates are shown as obtained from an OPES simulation using (c) 2-state and (d) 4-state DeepTDA CV$_c$. Points colored according to the corresponding CV$_c$ value. Configurations obtained from MoP simulations using (e) the initial 2-state and (f) the 4-state DeepTDA-based end-to-end distance CV$_t$ are shown. The violet and blue paths show the most probable reactive and trapped trajectories, respectively, i.e., the ones attaining the lowest value of the Onsager-Machlup action. In all panels, the isolines of the model potential are shown in grey as a reference.}
    \label{fig:mbpot}
\end{figure}
We first applied our iterative protocol to a particle moving in the two-dimensional model potential (adapted from M\"uller and Brown \cite{mullerLocationSaddlePoints1979}) shown by the isolines of figure~\ref{fig:mbpot}. The potential has three metastable states: an initial basin A and a final basin C  which we assume to be known beforehand, and an intermediate basin B. The relative positions of the three minima were designed to provide a scenario in which neither coordinate axis can resolve the transition states and drive the exploration of the whole free energy surface. Furthermore, in this case, a neural network CV simply trained to discriminate between the A and C is likely to fail, as demonstrated below.

\begin{figure}[!ht]
    \centering
    \includegraphics[width=\linewidth]{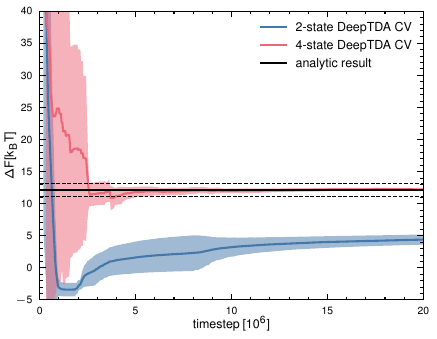}
    \caption{Free energy difference between the left and right basins of the model potential, as estimated from 5 independent OPES simulations biasing the 4-state DeepTDA CV, as a function of simulation time. The analytical result of $12.15\, k_BT$ was obtained by numerical integration over the MB potential~\cite{invernizziExplorationVsConvergence2022a} and is indicated in black, with dotted lines indicating a margin of $1\, k_BT$.}
    \label{fig:mbpot_fed}
\end{figure}

We first performed unbiased simulations in the A and C basins and trained an initial 2-state DeepTDA CV$_c$. The value of the latter is shown by the colored map reported in figure~\ref{fig:mbpot}a. Clearly, the intermediate metastable state B is not discriminated from the C basin since  the CV attains the same value in the two basins. As a consequence, when used in an OPES simulation, we found that this CV$_c$ is very inefficient in guiding transitions between A and C. Furthermore, during the simulation, the system is driven to sample unphysical trajectories that differ greatly from the minimum energy pathway (see figure~\ref{fig:mbpot}c). As a result, we also observe that the free energy difference between basins A and C is not accurately estimated when compared to the analytical result obtained by numerical integration of the potential (see figure~\ref{fig:mbpot_fed}).

Nevertheless, we can employ this suboptimal CV to construct the end-to-end distance CV$_t$ defined in equation~\ref{eq:traj_cv} for use in a MoP simulation. The samples obtained from the simulation in trajectory space are reported in figure~\ref{fig:mbpot}e. Notably, the sampled trajectories follow the underlying minimum energy pathways. This is due to the forces driving the polymer dynamics not being directly related to the potential energy surface but rather to the Onsager-Machlup action, which is lower for the more statistically relevant ones. By using the end-to-end distance CV$_t$, the metadynamics bias acts only on the polymer endpoints, while the intermediate replicas are free to relax, minimizing the OM action. This illustrates the robustness of MoP in sampling physically relevant trajectories even when using suboptimal CV$_t$. Importantly, the analysis of the data allows blindly detecting the intermediate state from the presence of crumpled polymers confined entirely into this basin, which are characterized by small values of the end-to-end distance CV$_t$, $S\sim0$. 

Partial reactive trajectories connecting the A and B basins were also observed (see figure~S4b). However, the simulation could not sample complete reactive paths connecting from A to C due to the suboptimal CV$_c$ used in equation~\eqref{eq:traj_cv}, which cannot distinguish correctly between B and C. We solve this problem by performing a second iteration of the algorithm in which the information gained from the MoP run is used to train a refined, 4-states DeepTDA CV, including data from the three metastable states plus the transition region between A and B (for all technical details we refer to the Materials and Methods). The colored map of the new CV$_c$ is shown in figure~\ref{fig:mbpot}b. It is apparent that all relevant metastable and transition states are resolved.

The new CV$_{c,t}$ drives complete  transitions from A to C both when used in MoP (figure~\ref{fig:mbpot}f) as well as in standard OPES simulations (figure~\ref{fig:mbpot}d). Figure~\ref{fig:mbpot_fed} shows that the free energy difference between basins A and C, as estimated with the new CV$_c$, is in excellent agreement with the analytical result. We also checked that the corresponding end-to-end distance CV$_{t}$ improves sampling in trajectory space. Figure~\ref{fig:mbpot}f reports the result of a MoP simulation, showing the sampling of complete reactive trajectories connecting A and C along the minimum free energy path. The efficiency of this CV is further demonstrated by the fact that it was able to generate also the partial paths connecting basins A and B, and B and C (see figure~S5b). From the complete reactive paths, we can also observe that they indeed spend an increased amount of time in the vicinity of the transition state, as illustrated in figure~S6. 

The new dataset obtained from MoP allowed us to train a 5-states DeepTDA CV$_{c,t}$, also including data from the transition state between B and C. The resulting CVs, however, did not lead to significant improvements respect to the 4-states versions.

\paragraph{Alanine Dipeptide}
\begin{figure}[!ht]
    \centering
    \includegraphics[width=\linewidth]{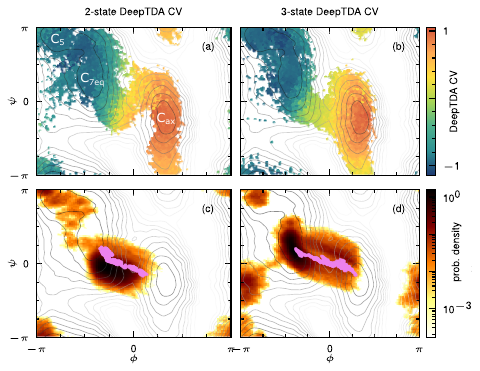}
    \caption{Biased simulations in configuration and trajectory space of alanine dipeptide. Scatter plots of $x,y$ coordinates in OPES simulation using (a) 2-state and (b) 3-state DeepTDA CV$_c$ are shown, colored according to CV value. MoP simulations using (c) initial and (d) 3-state DeepTDA CV$_t$ are shown, shown as normalized density of all configurations in logarithmic scale. The violet paths show the most probable reactive trajectories, i.e. the ones attaining the lowest value of the Onsager-Machlup action. In all panels, isolines of the FES obtained from a reference calculation are shown in grey.}
    \label{fig:alanine}
\end{figure}
We now move to the conformational dynamics of alanine dipeptide in vacuum. The free energy surface in the Ramachandran plane spanned by the  dihedral angles $\phi$ and $\psi$ is indicated by the grey isolines in figure~\ref{fig:alanine}. The system is characterized by the presence of three metastable states, labeled $C_5$, $C_{7eq}$ and $C_{ax}$. Specifically, the $C_5$ and $C_{7eq}$ conformers are separated by a barrier of the order of a few $k_BT$~\cite{vargasConformationalStudyAlanine2002a} and form a unique basin at room temperature, while a minimum barrier of around 13 $k_BT$ separates $C_{7eq}$ and $C_{ax}$. 

\begin{figure}[!ht]
    \centering
    \includegraphics[width=.45\textwidth]{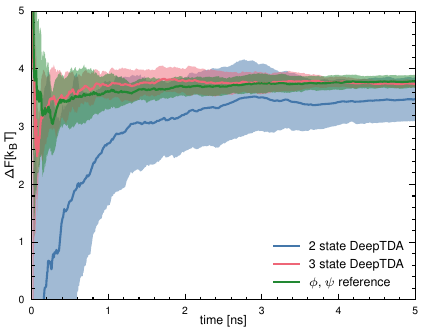}
    \caption{Free energy difference between the $C_{7eq}$ and $C_{ax}$ basins of alanine dipeptide over time. Shown are the results obtained from 5 independent enhanced sampling simulations respectively, biasing the 2-state and 3-state DeepTDA CV, as well as reference calculation using the dihedral angles $\phi,\psi$ as CV.}
    \label{fig:alanine_df}
\end{figure}

As done in the previous example, we start by performing unbiased simulations in the two metastable states, and training an initial 2-state DeepTDA CV$_c$. In agreement with Ref.~\cite{trizioEnhancedSamplingReaction2021}, we found that this CV is already able to drive transitions across the barrier separating $C_{7eq}$ and $C_{ax}$ to an acceptable degree when used in an OPES simulation (see figure~\ref{fig:alanine}a). However, in doing so, the system does not follow precisely the expected minimum free energy path but it samples also trajectories crossing high energy barriers around $\psi\sim\pi/2$. This can be explained by the lack of  transition state  data. ~\cite{trizioEnhancedSamplingReaction2021}. Figure~\ref{fig:alanine_df} reports a comparison of the performance of this CV$_c$ with the results obtained from a reference OPES simulation performed biasing the $\phi$ and $\psi$ angles, showing slow convergence and a significant discrepancy of $0.2\,k_BT$ in the free energy difference after 5~ns of simulation time. 

Figure~\ref{fig:alanine}c shows the result of MoP using the corresponding CV$_t$ in trajectory space, built following equation~\eqref{eq:traj_cv}. It can be seen that the sampling is focused on trajectories which follow the minimum free energy path. However, diffusion between the basins $C_5$ and $C_7eq$ is slow and only one transition takes place. This is due to the fact that these two minima are not distinguished by the DeepTDA CV$_c$. Furthermore, no paths reach fully into the $C_{ax}$ basin. Again, this can be explained by the lack of information on the transition state, which causes the starting CV to reach its maximum value before the true potential minimum is reached. Nonetheless, we can extract data from the transition state from the sampled reactive paths to train a new, 3-state DeepTDA CV$_c$ (see Materials and Methods for details). In figure~\ref{fig:alanine}b, we observe that in configuration space, transitions are now confined more closely to the minimum free energy paths, improving the efficiency of the CV$_c$. This is further supported by figure~\ref{fig:alanine_df}, demonstrating improved convergence speed and smaller statistical fluctuations of the 3-state DeepTDA CV$_c$, as compared to the 2-state DeepTDA CV$_c$, and similar convergence speed as the reference calculation. When used in a MoP simulation, reactive paths are sampled that reach considerably further inside the $C_{ax}$ state (see figure~\ref{fig:alanine}d) and fully connecting the $C_{7eq}$ and $C_{ax}$ basins. We also note that, after starting in the $C_5$ basin, the simulation takes the less likely, but known path \cite{mandelliModifiedNudgedElastic2021} across the higher barrier between $C_5$ and $C_{7eq}$, before sampling $C_{7eq}$ and reactive paths between $C_{7eq}$ and $C_{ax}$. However, these configurations are sampled during the initial part of the OPES-based MoP simulation – where the bias deposition is particularly aggressive – and are therefore exempted from further analysis.
\paragraph{Conclusions}
We have presented an iterative approach based on the metadynamics of paths algorithm~\cite{mandelliMetadynamicsPaths2020a} to reconstruct free energy landscapes as a function of data-driven CVs using datasets supplemented with configurations from the transition state ensemble. In doing so, we have also addressed directly the problem of designing efficient CVs in trajectory space. We found that the augmentation with MoP data leads to a significant performance increase of the learned CVs in configurational space. Good CVs could be generated even when using MoP in an exploratory manner (i.e., convergence was not needed), thus considerably reducing the computational effort required to obtain meaningful results.

Besides being the only path sampling method so far enabling the exploration of free energy landscapes using CVs, the use of MoP to obtain transition state data offers several other advantages: (i) in contrast to previous approaches~\cite{rayDeepLearningCollective2023,bestReactionCoordinatesRates2005,hooftDiscoveringCollectiveVariables2021,sunMultitaskMachineLearning2022}, MoP enables the sampling of (non)reactive trajectories in an unconstrained manner and is robust to the choice of sub-optimal CVs; (ii) running MoP amounts to a single metadynamics simulation, considerably simplifying the methodology compared to Monte Carlo approaches like transition path sampling; (iii) MoP can be implemented in an extremely parallel fashion~\cite{calhounHyperparallelAlgorithmsCentroid1996} to exploit modern massively parallel supercomputers (which have recently breached the exascale limit); (iv) compared to chain-of-states methods~\cite{eStringMethodStudy2002,jonssonNudgedElasticBand1998,mandelliModifiedNudgedElastic2021} MoP generates a large amount of configurations, which is required for data-driven applications like ours.

The efficiency of our procedure was tested in two models of increasing complexity, including a simple molecular system. We expect our protocol to aid in the discovery of collective variables, the exploration of transition pathways, and the estimation of free energy profiles in complex systems of relevance in biology, chemistry, and materials science. Future work will focus on scaling the proposed protocol to more challenging, real-world applications. While MoP can be readily applied to the condensed phase, including explicit solvent, applications to large macromolecules of our ML approach will need a suitable set of input descriptors such as backbone dihedral angles, per-residue Q-values, or water coordination numbers~\cite{vaniAlphaFold2RAVESequenceBoltzmann2023,rizziRoleWaterHostguest2021}.

Finally, we note that, in this work, we have focused on building ML CVs that use only the spatial distribution of the samples obtained from MoP data. However, learning dynamical information from the trajectory data~\cite{perez-hernandezIdentificationSlowMolecular2013a,wehmeyerTimelaggedAutoencodersDeep2018,bonatiDeepLearningSlow2021} is another very appealing avenue, as the sampled paths carry information on the unbiased dynamics of the system.

\section*{Materials and Methods}
\paragraph{DeepTDA CVs}
The DeepTDA neural network CVs~\cite{trizioEnhancedSamplingReaction2021} are trained in PyTorch using an existing implementation in the mlcolvar package~\cite{bonatiUnifiedFrameworkMachine2023a}, in its release v0.2.0. The networks use a feed-forward architecture with 3 hidden layers with $\{24,12,1\}$ nodes and the ReLU activation function.
To enforce the target distribution on the latent CV space, the objective function for the neural network is chosen to be a mean-squared error between the mean and standard deviation at the hidden layer and their respective target values, $\mu^{tg}$ and $\sigma^{tg}$, which in the one-dimensional case is given by,
\begin{equation}
    L = \sum_k^{N_s} \alpha (\mu_{k} - \mu_{k}^{tg})^2 + \beta (\sigma_{k} - \sigma_{k}^{tg})^2,
    \label{eq:tdaloss}
\end{equation}
where $k$ denotes the $N_s$ different states, and the hyperparameters $\alpha$ and $\beta$ ensuring adequate scaling of the respective loss terms. The resulting DeepTDA CV is normalized over the training data to a range of $-1 \leq s \leq 1$. \\
In all trained networks, the parameters are optimized using the ADAM optimizer~\cite{kingmaAdamMethodStochastic2017} with a learning rate of $10^{-3}$. In addition to the TDA loss in equation \ref{eq:tdaloss}, L2 regularization with $\lambda=10^{-5}$ has been added to the weights. The training is stopped when convergence of the loss function is reached, using an Early-Stopping routine with patience set to 15 epochs to avoid overfitting. The hyperparameters $\alpha$ and $\beta$ were set to values of $1$ and $250$, respectively. 

\paragraph{Two-Dimensional Model Potential}
To evaluate the proposed methodology, we designed a two-dimensional model potential, based on the analytical form introduced by Müller and Brown \cite{mullerLocationSaddlePoints1979}. The isolines of this modified potential are shown in figure~S1, and has the analytical from 
\begin{align}\begin{split}
    U_{\text{MB}}(x,y) &= \sum_k A_k \exp\left[ a_k(x-x_k^0)^2 \right. \\ 
        &\left.\quad + b_k(x-x_k^0)(y-y_k^0) + c_k(y-y_k^0)^2 \right], \\
    \vb A &= (-16, -11, -17, 2), \\
    \vb a &= (-10, -1, -6.5, 0.4), \\
    \vb b &= (5, 0, 11, 0), \\
    \vb c &= (-5, -10, -6.5, 1.1), \\
    \vb x_0 &= (1, 0, -0.5, 0), \\
    \vb y_0 &= (1.2, 0.5, 1.5, 1).
    \label{eq:mbpot}
\end{split}\end{align}
All simulations in this work are started in basin A, where the potential assumes its global minimum. In this work, we consider Langevin dynamics of a single particle moving along the model potential. The simulations use natural units, such that $k_B=1$, and a temperature of $T=0.1$, placing the highest free energy barrier at around $120\, k_BT$. \\
Because the analytical form of the potential is known, the difference in free energy between the basins A and C can be calculated directly via numerical integration, with
\begin{equation}
    \Delta F = -\frac1\beta \log \frac{\int_C e^{-\beta U_{\text{MB}}(x,y)}dx\,dy}{\int_A e^{-\beta U_{\text{MB}}(x,y)}dx\,dy}
\end{equation}
where $A=\{(x,y) \mid y > x+1.5\}$ and $C=\{(x,y) \mid x-1.5 < y < x+1.5\}$, which in this case equates to $12.15 \, k_BT$.

The simulations in configurational space of the 2D model potential were performed using the molecular simulation engine LAMMPS \cite{thompsonLAMMPSFlexibleSimulation2022} in its stable release version from 23. June 2022, patched with PLUMED 2.8 \cite{tribelloPLUMEDNewFeathers2014}, which also provides an interface with the LibTorch C++ library to implement the neural network-based DeepTDA CVs \cite{bonatiUnifiedFrameworkMachine2023a}. The damping constant in the Langevin thermostat is set to a value of $0.1$, which corresponds to a value of $\nu=10$, and the time step is set to $\Delta t=0.01$ in arbitrary units of time. \\
To perform enhanced sampling simulations we here use the recently introduced OPES method \cite{invernizziRethinkingMetadynamicsBias2020a}, an implementation for which is provided in PLUMED. For detailed information about the usage of this bias potential and a definition of the relevant parameters, the reader is referred to the PLUMED documentation. For the OPES simulation with the initial 2-state DeepTDA CV, the  parameters are set to BARRIER~$=17$, PACE~$=300$ and SIGMA~$=0.03$. The SIGMA is chosen to reflect the standard deviation of the CV values in a short, unbiased simulation. For OPES simulation using the 4-state DeepTDA CV, the parameter values BARRIER~$=15$, PACE~$=300$ and SIGMA~$=0.01$. The SIGMA parameter is again chosen as the CVs standard deviation in unbiased simulation, and the BARRIER parameter is lowered to allow for faster convergence.

To carry out the Metadynamics of Paths simulations in trajectory space, an extra fix is added to the LAMMPS suite, adapted from Ref. \cite{mandelliMetadynamicsPaths2020a} to directly evaluate the analytical derivatives of the potential. The friction and time step parameter of the polymer are set to $\nu = 10$ and $\Delta t = 0.01$, respectively. 
The polymer length should be chosen to allow sampling the reactive trajectory of interest. Here, a value of N=288 was found to be sufficient. Note that thanks to its parallel implementation, the computational cost per time step of the MoP algorithm only weakly depends on the polymer size (see figure S2).
In trajectory space, the polymers are then sampled at a time step of $0.01$ and a damping constant of damp $= 100$. The initial configuration for the polymer was obtained by running an unbiased MetaD of Paths simulation starting from the same configuration for every bead at $\vb r = (-0.5, 1.5)^\intercal$ and letting the polymer relax for $10^4$ steps. While relaxing the polymer removes the dependence on the precise starting conformation, we found that it is necessary to run a sufficient number of relaxation steps in order to achieve a faster exploration of relevant reactive trajectories. \\
To enhance the sampling of the polymers, OPES is again used. The underlying configurational DeepTDA CV is first evaluated on each bead. Then, a modified version of the CUSTOM/MATHEVAL action is used to access the values from different beads, to evaluate the end-to-end difference for use as the trajectory CV. MoP simulations with both the initial DeepTDA CV and the 4-state CV are performed with OPES parameters set to BARRIER~$= 20$ and PACE~$= 500$, and an adaptive SIGMA. In both cases, the MoP simulations were run for $2\times 10^6$ steps, saving polymer configurations every 500 steps.

\paragraph{DeepTDA CV Training for 2D Model Potential}
For the initial DeepTDA CV $s_0$ on the unbiased training data, the target centers and widths are chosen as $\mu^{tg}=[-7,7]$ and $\sigma^{tg}=[0.2,0.2]$, respectively. The training data set consists of 6000 configurations from each of the initial metastable states, as shown labeled 'unbiased data' in figure~S3b. \\
We now train the 4-state DeepTDA CV in a two-step process, the reason for which is explained below. From the trajectory data, shown in Fig. \ref{fig:mbpot}e, we first isolate the reactive trajectories and those trapped in metastable states. This can be quantified by using the trajectory CV $S(\{\vb R_i\})$ in Eq. \ref{eq:traj_cv}, and we here classify as kinetically trapped or confined trajectories with values of $S \leq 0.3$. In this way, we find trajectories trapped in the initial basin A, but also those confined to the previously 'unknown' basin B. To separate these configurations between known and unknown states, OPTICS clustering~\cite{ankerstOPTICSOrderingPoints1999a} is used. \\
To filter the reactive trajectories, the range of values to select for depends strongly on the respective system under consideration, and how well the initial DeepTDA CV is able to discern different regions of the phase space, which is why we first train a 3-state DeepTDA CV $s_1$. 
For the training of this CV, target centers and widths for this CV are chosen as $\mu^{tg}=[-15,0,15]$ and $\sigma^{tg}=[0.3,1.0,0.3]$ respectively, corresponding to a respective separation of $\Delta=15$. Evaluating this CV on the previously obtained trajectory data, we can now much more confidently set a threshold of $\abs{S\{\vb R_i\}} \geq 1$ to select trajectories that connect A and B. 
Now, information from the reactive paths is taken into account, by selecting configurations that satisfy $\mu_A+3\sigma_A \leq s_1(\vb r) \leq \mu_{int}-3\sigma_{int}$, where the $\mu_{A/int}$ and $\sigma_{A/int}$ denote mean and standard deviation in the initial/intermediate metastable states, respectively. The precise multiple of the standard deviation should be chosen carefully, as to select data that covers as wide a CV$_c$ range as possible, without introducing overlaps. This data is added as a fourth state for the training of the 4-state DeepTDA CV $s_2$, using the target widths $\mu^{tg}=[-30,-15,0,15]$ and centers $\sigma^{tg}=[0.3,4.0,1.0,0.3]$, and the training data is shown in figure~S3a. These values are chosen by visual inspection of the CV histograms in the training data, to select for sharp peaks in the metastable states and broader, slightly overlapping distributions for the reactive paths and transition states (see figure S3b). In both cases, a random subset of 6000 configurations is selected from the trapped and reactive trajectory data, to match the amount of unbiased data.

\paragraph{Alanine Dipeptide}
For the simulations of the conformational dynamics of alanine dipeptide in vacuum, we again used LAMMPS patched with PLUMED. The Amber99-SB~\cite{hornakComparisonMultipleAmber2006} force field was used, converted for use in LAMMPS using the convert.py script in the InterMol~\cite{shirtsLessonsLearnedComparing2017} software. We consider Langevin dynamics in an NVE ensemble at a temperature of 300 K, and use a time step of 0.5 fs and a dampening constant of 500 fs. 
To perform biased simulations with the DeepTDA CVs in OPES, we used parameter values BARRIER~$=50$, PACE~$=100$, and an adaptive SIGMA.

To carry out MoP simulations, we used the path\_dynamics fix provided by and described in Ref.~\cite{mandelliMetadynamicsPaths2020a}. The parameters of the polymer are set to $\nu = 0.25 \frac1s$ and $\Delta t = 1 fs$. In trajectory space, the polymers, made up of $N = 512$ beads, are then sampled at a time step of $1.0$ and a damping constant of damp~$= 1000$ in trajectory space. The initial configuration for the polymer was obtained by running an unbiased simulation in configuration space for $10^6$ steps, and saving the last 512 steps as initial configurations for the polymer beads. The polymer was then allowed to equilibrate in an unbiased MoP simulation for $10^6$ steps. As described above, we again use OPES to drive the sampling of the polymers with a bias potential, using parameter values BARRIER~$=80$ and PACE~$=1000$, as well as a bias factor of 15. We again ran MoP simulations for $4\times 10^6$ steps for both realizations of the DeepTDA CV, saving polymer configurations every 1000 steps.

The DeepTDA neural network-CVs for Alanine Dipeptide are trained with largely identical settings to the case of the model potential, with the exception of a learning rate of $10^{-3}$. As a set of descriptors, we use the set of pairwise distances between the heavy atoms of alanine dipeptide, as compiled by Bonati and coworkers~\cite{bonatiDataDrivenCollectiveVariables2020}, thereby insuring rototranslational invariance of our CV. \\
The 2-state DeepTDA CV $s_0$ is trained with 4000 configurations each from the $C_5/C_{7eq}$ and $C_{ax}$ basins, respectively. As values for the target centers and widths for the modes in CV spaces, we again choose $\mu^{tg}=[-7,7]$ and $\sigma^{tg}=[0.2,0.2]$. Proceeding as described in the main text, by first separating the trajectory data, sampled with MoP using the 2-state CV, into trapped paths with a value of $S(\{\vb R_i\}) \leq 0.5$ and reactive paths with $S(\{\vb R_i\}) \geq 1.2$. In this case, we don't observe any confined data in previously unexplored regions, so we proceed by integrating data from the transition state as an intermediary. The training data for the 3-state DeepTDA CV $s_1$ is thus chosen by selecting configurations with $s_0 \geq 0.3$ from the reactive paths, as shown in figure~S11. From this data, we again choose a random subset of 4000 configurations to match the amount of unbiased data, and train the new CV by choosing target centers and widths $\mu^{tg}=[-7,0,7]$ and $\sigma^{tg}=[0.3,1.0,0.3]$. 

\paragraph{Data Availability}
The  input files for LAMMPS and PLUMED used to generate the presented results, scripts to compile the necessary software, as well as example code to train the DeepTDA NN CVs, can be found on GitHub at \url{https://github.com/lmuellender/MoP-DeepTDA}. Additionally, all data and files are available via Zenodo at \url{https://doi.org/10.5281/zenodo.10845989}.

\paragraph{Acknowledgement}
This research was partly supported by European Union’s HORIZON MSCA Doctoral Networks programme, under Grant Agreement No. 101072344, project AQTIVATE (Advanced computing, QuanTum algorIthms and data-driVen Approaches for science, Technology and Engineering).
Furhtermore, the authors acknowledge support from the Helmholtz European Partnering program (“Innovative high-performance computing approaches for molecular neuromedicine”) and the Swedish eScience Research Center.
The authors also gratefully acknowledge the Gauss Centre for Supercomputing e.V. (\url{www.gauss-centre.eu}) for funding this project by providing computing time through the John von Neumann Institute for Computing (NIC) on the GCS Supercomputer JUWELS~\cite{alvarezJUWELSClusterBooster2021} at J\"{u}lich Supercomputing Centre (JSC).
This article was posted to a preprint: \url{https://doi.org/10.48550/arXiv.2311.05571}

\printbibliography[heading=bibintoc, title={References}]

@article{rizziRoleWaterHostguest2021,
  title = {The Role of Water in Host-Guest Interaction},
  author = {Rizzi, Valerio and Bonati, Luigi and Ansari, Narjes and Parrinello, Michele},
  year = {2021},
  month = jan,
  journal = {Nature Communications},
  volume = {12},
  number = {1},
  pages = {93},
  publisher = {{Nature Publishing Group}},
  doi = {10.1038/s41467-020-20310-0},
  copyright = {2021 The Author(s)}
}

@article{eStringMethodStudy2002,
  title = {String Method for the Study of Rare Events},
  author = {E, Weinan and Ren, Weiqing and {Vanden-Eijnden}, Eric},
  year = {2002},
  month = aug,
  journal = {Physical Review B},
  volume = {66},
  number = {5},
  pages = {052301},
  publisher = {{American Physical Society}},
  doi = {10.1103/PhysRevB.66.052301}
}

@article{vaniAlphaFold2RAVESequenceBoltzmann2023,
  title = {{{AlphaFold2-RAVE}}: {{From Sequence}} to {{Boltzmann Ranking}}},
  shorttitle = {{{AlphaFold2-RAVE}}},
  author = {Vani, Bodhi P. and Aranganathan, Akashnathan and Wang, Dedi and Tiwary, Pratyush},
  year = {2023},
  month = jul,
  journal = {Journal of Chemical Theory and Computation},
  volume = {19},
  number = {14},
  pages = {4351--4354},
  doi = {10.1021/acs.jctc.3c00290}
}

@article{alvarezJUWELSClusterBooster2021,
  title = {{{JUWELS Cluster}} and {{Booster}}: {{Exascale Pathfinder}} with {{Modular Supercomputing Architecture}} at {{Juelich Supercomputing Centre}}},
  shorttitle = {{{JUWELS Cluster}} and {{Booster}}},
  author = {Alvarez, Damian},
  year = {2021},
  month = oct,
  journal = {Journal of large-scale research facilities JLSRF},
  volume = {7},
  pages = {A183-A183},
  doi = {10.17815/jlsrf-7-183},
  copyright = {Copyright (c) 2021 Journal of large-scale research facilities JLSRF}
}

@article{ankerstOPTICSOrderingPoints1999a,
  title = {{{OPTICS}}: Ordering Points to Identify the Clustering Structure},
  shorttitle = {{{OPTICS}}},
  author = {Ankerst, Mihael and Breunig, Markus M. and Kriegel, Hans-Peter and Sander, J{\"o}rg},
  year = {1999},
  month = jun,
  journal = {ACM SIGMOD Record},
  volume = {28},
  number = {2},
  pages = {49--60},
  doi = {10.1145/304181.304187}
}

@article{barducciWellTemperedMetadynamicsSmoothly2008b,
  title = {Well-{{Tempered Metadynamics}}: {{A Smoothly Converging}} and {{Tunable Free-Energy Method}}},
  shorttitle = {Well-{{Tempered Metadynamics}}},
  author = {Barducci, Alessandro and Bussi, Giovanni and Parrinello, Michele},
  year = {2008},
  month = jan,
  journal = {Physical Review Letters},
  volume = {100},
  number = {2},
  pages = {020603},
  publisher = {{American Physical Society}},
  doi = {10.1103/PhysRevLett.100.020603}
}

@article{bestReactionCoordinatesRates2005,
  title = {Reaction Coordinates and Rates from Transition Paths},
  author = {Best, Robert B. and Hummer, Gerhard},
  year = {2005},
  month = may,
  journal = {Proceedings of the National Academy of Sciences},
  volume = {102},
  number = {19},
  pages = {6732--6737},
  doi = {10.1073/pnas.0408098102}
}

@article{blaakCrystalNucleationColloidal2004,
  title = {Crystal {{Nucleation}} of {{Colloidal Suspensions}} under {{Shear}}},
  author = {Blaak, Ronald and Auer, Stefan and Frenkel, Daan and L{\"o}wen, Hartmut},
  year = {2004},
  month = aug,
  journal = {Physical Review Letters},
  volume = {93},
  number = {6},
  pages = {068303},
  doi = {10.1103/PhysRevLett.93.068303}
}

@article{bolhuisTransitionPathSampling2002,
  title = {Transition {{Path Sampling}}: {{Throwing Ropes Over Rough Mountain Passes}}, in the {{Dark}}},
  shorttitle = {{{TRANSITION PATH SAMPLING}}},
  author = {Bolhuis, Peter G. and Chandler, David and Dellago, Christoph and Geissler, Phillip L.},
  year = {2002},
  journal = {Annual Review of Physical Chemistry},
  volume = {53},
  number = {1},
  pages = {291--318},
  doi = {10.1146/annurev.physchem.53.082301.113146},
  pmid = {11972010}
}

@article{bonatiDataDrivenCollectiveVariables2020,
  title = {Data-{{Driven Collective Variables}} for {{Enhanced Sampling}}},
  author = {Bonati, Luigi and Rizzi, Valerio and Parrinello, Michele},
  year = {2020},
  month = apr,
  journal = {The Journal of Physical Chemistry Letters},
  volume = {11},
  number = {8},
  pages = {2998--3004},
  doi = {10.1021/acs.jpclett.0c00535}
}

@article{bonatiDeepLearningSlow2021,
  title = {Deep Learning the Slow Modes for Rare Events Sampling},
  author = {Bonati, Luigi and Piccini, GiovanniMaria and Parrinello, Michele},
  year = {2021},
  month = nov,
  journal = {Proceedings of the National Academy of Sciences},
  volume = {118},
  number = {44},
  pages = {e2113533118},
  doi = {10.1073/pnas.2113533118}
}

@article{bonatiUnifiedFrameworkMachine2023a,
  title = {A Unified Framework for Machine Learning Collective Variables for Enhanced Sampling Simulations: Mlcolvar},
  shorttitle = {A Unified Framework for Machine Learning Collective Variables for Enhanced Sampling Simulations},
  author = {Bonati, Luigi and Trizio, Enrico and Rizzi, Andrea and Parrinello, Michele},
  year = {2023},
  month = jul,
  journal = {The Journal of Chemical Physics},
  volume = {159},
  number = {1},
  pages = {014801},
  doi = {10.1063/5.0156343}
}

@article{calhounHyperparallelAlgorithmsCentroid1996,
  title = {Hyper-Parallel Algorithms for Centroid Molecular Dynamics: Application to Liquid Para-Hydrogen},
  shorttitle = {Hyper-Parallel Algorithms for Centroid Molecular Dynamics},
  author = {Calhoun, August and Pavese, Marc and Voth, Gregory A.},
  year = {1996},
  month = nov,
  journal = {Chemical Physics Letters},
  volume = {262},
  number = {3-4},
  pages = {415--420},
  doi = {10.1016/0009-2614(96)01109-8}
}

@article{chenCollectiveVariablebasedEnhanced2021,
  title = {Collective Variable-Based Enhanced Sampling and Machine Learning},
  author = {Chen, Ming},
  year = {2021},
  month = oct,
  journal = {The European Physical Journal B},
  volume = {94},
  number = {10},
  pages = {211},
  doi = {10.1140/epjb/s10051-021-00220-w}
}

@article{chenMolecularEnhancedSampling2018,
  title = {Molecular Enhanced Sampling with Autoencoders: {{On-the-fly}} Collective Variable Discovery and Accelerated Free Energy Landscape Exploration},
  shorttitle = {Molecular Enhanced Sampling with Autoencoders},
  author = {Chen, Wei and Ferguson, Andrew L.},
  year = {2018},
  journal = {Journal of Computational Chemistry},
  volume = {39},
  number = {25},
  pages = {2079--2102},
  doi = {10.1002/jcc.25520}
}

@article{darveAdaptiveBiasingForce2008,
  title = {Adaptive Biasing Force Method for Scalar and Vector Free Energy Calculations},
  author = {Darve, Eric and {Rodr{\'i}guez-G{\'o}mez}, David and Pohorille, Andrew},
  year = {2008},
  month = apr,
  journal = {The Journal of Chemical Physics},
  volume = {128},
  number = {14},
  pages = {144120},
  doi = {10.1063/1.2829861}
}

@article{delemotteFreeenergyLandscapeIonchannel2015,
  title = {Free-Energy Landscape of Ion-Channel Voltage-Sensor\textendash Domain Activation},
  author = {Delemotte, Lucie and Kasimova, Marina A. and Klein, Michael L. and Tarek, Mounir and Carnevale, Vincenzo},
  year = {2015},
  month = jan,
  journal = {Proceedings of the National Academy of Sciences},
  volume = {112},
  number = {1},
  pages = {124--129},
  publisher = {{Proceedings of the National Academy of Sciences}},
  doi = {10.1073/pnas.1416959112}
}

@article{dellagoTransitionPathSampling1998,
  title = {Transition Path Sampling and the Calculation of Rate Constants},
  author = {Dellago, Christoph and Bolhuis, Peter G. and Csajka, F{\'e}lix S. and Chandler, David},
  year = {1998},
  month = feb,
  journal = {The Journal of Chemical Physics},
  volume = {108},
  number = {5},
  pages = {1964--1977},
  doi = {10.1063/1.475562}
}

@article{heninEnhancedSamplingMethods2022a,
  title = {Enhanced {{Sampling Methods}} for {{Molecular Dynamics Simulations}} [{{Article}} v1.0]},
  author = {H{\'e}nin, J{\'e}r{\^o}me and Leli{\`e}vre, Tony and Shirts, Michael R. and Valsson, Omar and Delemotte, Lucie},
  year = {2022},
  month = dec,
  journal = {Living Journal of Computational Molecular Science},
  volume = {4},
  number = {1},
  pages = {1583--1583},
  doi = {10.33011/livecoms.4.1.1583},
  copyright = {Copyright (c) 2022 J\'er\^ome  H\'enin, Tony Leli\`evre, Michael Shirts, Omar Valsson, Lucie Delemotte}
}

@article{hooftDiscoveringCollectiveVariables2021,
  title = {Discovering {{Collective Variables}} of {{Molecular Transitions}} via {{Genetic Algorithms}} and {{Neural Networks}}},
  author = {Hooft, Ferry and {P{\'e}rez de Alba Ort{\'i}z}, Alberto and Ensing, Bernd},
  year = {2021},
  month = apr,
  journal = {Journal of Chemical Theory and Computation},
  volume = {17},
  number = {4},
  pages = {2294--2306},
  doi = {10.1021/acs.jctc.0c00981},
  pmcid = {PMC8047796},
  pmid = {33662202}
}

@article{hornakComparisonMultipleAmber2006,
  title = {Comparison of Multiple {{Amber}} Force Fields and Development of Improved Protein Backbone Parameters},
  author = {Hornak, Viktor and Abel, Robert and Okur, Asim and Strockbine, Bentley and Roitberg, Adrian and Simmerling, Carlos},
  year = {2006},
  journal = {Proteins: Structure, Function, and Bioinformatics},
  volume = {65},
  number = {3},
  pages = {712--725},
  doi = {10.1002/prot.21123},
  copyright = {Copyright \textcopyright{} 2006 Wiley-Liss, Inc.}
}

@article{invernizziExplorationVsConvergence2022a,
  title = {Exploration vs {{Convergence Speed}} in {{Adaptive-Bias Enhanced Sampling}}},
  author = {Invernizzi, Michele and Parrinello, Michele},
  year = {2022},
  month = jun,
  journal = {Journal of Chemical Theory and Computation},
  volume = {18},
  number = {6},
  pages = {3988--3996},
  publisher = {{American Chemical Society}},
  doi = {10.1021/acs.jctc.2c00152}
}

@article{invernizziRethinkingMetadynamicsBias2020a,
  title = {Rethinking {{Metadynamics}}: {{From Bias Potentials}} to {{Probability Distributions}}},
  shorttitle = {Rethinking {{Metadynamics}}},
  author = {Invernizzi, Michele and Parrinello, Michele},
  year = {2020},
  month = apr,
  journal = {The Journal of Physical Chemistry Letters},
  volume = {11},
  number = {7},
  pages = {2731--2736},
  publisher = {{American Chemical Society}},
  doi = {10.1021/acs.jpclett.0c00497}
}

@inproceedings{jonssonNudgedElasticBand1998,
  title = {Nudged Elastic Band Method for Finding Minimum Energy Paths of Transitions},
  booktitle = {Classical and {{Quantum Dynamics}} in {{Condensed Phase Simulations}}},
  author = {J{\'o}nsson, Hannes and Mills, Greg and Jacobsen, Karsten W.},
  year = {1998},
  month = jun,
  pages = {385--404},
  publisher = {{WORLD SCIENTIFIC}},
  address = {{LERICI, Villa Marigola}},
  doi = {10.1142/9789812839664_0016},
  isbn = {978-981-02-3498-0 978-981-283-966-4}
}

@misc{kingmaAdamMethodStochastic2017,
  title = {Adam: {{A Method}} for {{Stochastic Optimization}}},
  shorttitle = {Adam},
  author = {Kingma, Diederik P. and Ba, Jimmy},
  year = {2017},
  month = jan,
  number = {arXiv:1412.6980},
  eprint = {1412.6980},
  publisher = {{arXiv}},
  archiveprefix = {arxiv}
}

@article{laioEscapingFreeenergyMinima2002,
  title = {Escaping Free-Energy Minima},
  author = {Laio, Alessandro and Parrinello, Michele},
  year = {2002},
  month = oct,
  journal = {Proceedings of the National Academy of Sciences},
  volume = {99},
  number = {20},
  pages = {12562--12566},
  publisher = {{Proceedings of the National Academy of Sciences}},
  doi = {10.1073/pnas.202427399}
}

@article{lindorff-larsenHowFastFoldingProteins2011,
  title = {How {{Fast-Folding Proteins Fold}}},
  author = {{Lindorff-Larsen}, K. and Piana, S. and Dror, R. O. and Shaw, D. E.},
  year = {2011},
  month = oct,
  journal = {Science},
  volume = {334},
  number = {6055},
  pages = {517--520},
  doi = {10.1126/science.1208351}
}

@article{maAutomaticMethodIdentifying2005,
  title = {Automatic {{Method}} for {{Identifying Reaction Coordinates}} in {{Complex Systems}}},
  author = {Ma, Ao and Dinner, Aaron R.},
  year = {2005},
  month = apr,
  journal = {The Journal of Physical Chemistry B},
  volume = {109},
  number = {14},
  pages = {6769--6779},
  doi = {10.1021/jp045546c}
}

@article{mandelliMetadynamicsPaths2020a,
  title = {Metadynamics of {{Paths}}},
  author = {Mandelli, Davide and Hirshberg, Barak and Parrinello, Michele},
  year = {2020},
  month = jul,
  journal = {Physical Review Letters},
  volume = {125},
  number = {2},
  pages = {026001},
  publisher = {{American Physical Society}},
  doi = {10.1103/PhysRevLett.125.026001}
}

@article{mandelliModifiedNudgedElastic2021,
  title = {A Modified Nudged Elastic Band Algorithm with Adaptive Spring Lengths},
  author = {Mandelli, D. and Parrinello, M.},
  year = {2021},
  month = aug,
  journal = {The Journal of Chemical Physics},
  volume = {155},
  number = {7},
  pages = {074103},
  doi = {10.1063/5.0059593}
}

@article{mendelsCollectiveVariablesLocal2018,
  title = {Collective {{Variables}} from {{Local Fluctuations}}},
  author = {Mendels, Dan and Piccini, GiovanniMaria and Parrinello, Michele},
  year = {2018},
  month = jun,
  journal = {The Journal of Physical Chemistry Letters},
  volume = {9},
  number = {11},
  pages = {2776--2781},
  doi = {10.1021/acs.jpclett.8b00733}
}

@article{mullerLocationSaddlePoints1979,
  title = {Location of Saddle Points and Minimum Energy Paths by a Constrained Simplex Optimization Procedure},
  author = {M{\"u}ller, Klaus and Brown, Leo D.},
  year = {1979},
  journal = {Theoretica Chimica Acta},
  volume = {53},
  number = {1},
  pages = {75--93},
  doi = {10.1007/BF00547608}
}

@article{onsagerFluctuationsIrreversibleProcesses1953,
  title = {Fluctuations and {{Irreversible Processes}}},
  author = {Onsager, L. and Machlup, S.},
  year = {1953},
  month = sep,
  journal = {Physical Review},
  volume = {91},
  number = {6},
  pages = {1505--1512},
  doi = {10.1103/PhysRev.91.1505}
}

@article{perez-hernandezIdentificationSlowMolecular2013a,
  title = {Identification of Slow Molecular Order Parameters for {{Markov}} Model Construction},
  author = {{P{\'e}rez-Hern{\'a}ndez}, Guillermo and Paul, Fabian and Giorgino, Toni and De Fabritiis, Gianni and No{\'e}, Frank},
  year = {2013},
  month = jul,
  journal = {The Journal of Chemical Physics},
  volume = {139},
  number = {1},
  pages = {015102},
  doi = {10.1063/1.4811489}
}

@article{prattStatisticalMethodIdentifying1986,
  title = {A Statistical Method for Identifying Transition States in High Dimensional Problems},
  author = {Pratt, Lawrence R.},
  year = {1986},
  month = nov,
  journal = {The Journal of Chemical Physics},
  volume = {85},
  number = {9},
  pages = {5045--5048},
  doi = {10.1063/1.451695}
}

@article{rayDeepLearningCollective2023,
  title = {Deep Learning Collective Variables from Transition Path Ensemble},
  author = {Ray, Dhiman and Trizio, Enrico and Parrinello, Michele},
  year = {2023},
  month = may,
  journal = {The Journal of Chemical Physics},
  volume = {158},
  number = {20},
  pages = {204102},
  doi = {10.1063/5.0148872}
}

@article{ribeiroReweightedAutoencodedVariational2018,
  title = {Reweighted Autoencoded Variational {{Bayes}} for Enhanced Sampling ({{RAVE}})},
  author = {Ribeiro, Jo{\~a}o Marcelo Lamim and Bravo, Pablo and Wang, Yihang and Tiwary, Pratyush},
  year = {2018},
  month = aug,
  journal = {The Journal of Chemical Physics},
  volume = {149},
  number = {7},
  pages = {072301},
  doi = {10.1063/1.5025487}
}

@article{rohrdanzDiscoveringMountainPasses2013,
  title = {Discovering {{Mountain Passes}} via {{Torchlight}}: {{Methods}} for the {{Definition}} of {{Reaction Coordinates}} and {{Pathways}} in {{Complex Macromolecular Reactions}}},
  shorttitle = {Discovering {{Mountain Passes}} via {{Torchlight}}},
  author = {Rohrdanz, Mary A. and Zheng, Wenwei and Clementi, Cecilia},
  year = {2013},
  journal = {Annual Review of Physical Chemistry},
  volume = {64},
  number = {1},
  pages = {295--316},
  doi = {10.1146/annurev-physchem-040412-110006},
  pmid = {23298245}
}

@article{shirtsLessonsLearnedComparing2017,
  title = {Lessons Learned from Comparing Molecular Dynamics Engines on the {{SAMPL5}} Dataset},
  author = {Shirts, Michael R. and Klein, Christoph and Swails, Jason M. and Yin, Jian and Gilson, Michael K. and Mobley, David L. and Case, David A. and Zhong, Ellen D.},
  year = {2017},
  month = jan,
  journal = {Journal of Computer-Aided Molecular Design},
  volume = {31},
  number = {1},
  pages = {147--161},
  doi = {10.1007/s10822-016-9977-1}
}

@article{sormaniExplicitCharacterizationFreeEnergy2020b,
  title = {Explicit {{Characterization}} of the {{Free-Energy Landscape}} of a {{Protein}} in the {{Space}} of {{All Its C}} {\textsubscript{{$\alpha$}}} {{Carbons}}},
  author = {Sormani, Giulia and Rodriguez, Alex and Laio, Alessandro},
  year = {2020},
  month = jan,
  journal = {Journal of Chemical Theory and Computation},
  volume = {16},
  number = {1},
  pages = {80--87},
  doi = {10.1021/acs.jctc.9b00800}
}

@article{sultanAutomatedDesignCollective2018,
  title = {Automated Design of Collective Variables Using Supervised Machine Learning},
  author = {Sultan, Mohammad M. and Pande, Vijay S.},
  year = {2018},
  month = sep,
  journal = {The Journal of Chemical Physics},
  volume = {149},
  number = {9},
  pages = {094106},
  doi = {10.1063/1.5029972}
}

@article{sunMultitaskMachineLearning2022,
  title = {Multitask {{Machine Learning}} of {{Collective Variables}} for {{Enhanced Sampling}} of {{Rare Events}}},
  author = {Sun, Lixin and Vandermause, Jonathan and Batzner, Simon and Xie, Yu and Clark, David and Chen, Wei and Kozinsky, Boris},
  year = {2022},
  month = apr,
  journal = {Journal of Chemical Theory and Computation},
  volume = {18},
  number = {4},
  pages = {2341--2353},
  doi = {10.1021/acs.jctc.1c00143}
}

@article{thompsonLAMMPSFlexibleSimulation2022,
  title = {{{LAMMPS}} - a Flexible Simulation Tool for Particle-Based Materials Modeling at the Atomic, Meso, and Continuum Scales},
  author = {Thompson, Aidan P. and Aktulga, H. Metin and Berger, Richard and Bolintineanu, Dan S. and Brown, W. Michael and Crozier, Paul S. and {in 't Veld}, Pieter J. and Kohlmeyer, Axel and Moore, Stan G. and Nguyen, Trung Dac and Shan, Ray and Stevens, Mark J. and Tranchida, Julien and Trott, Christian and Plimpton, Steven J.},
  year = {2022},
  month = feb,
  journal = {Computer Physics Communications},
  volume = {271},
  pages = {108171},
  doi = {10.1016/j.cpc.2021.108171}
}

@article{tiwaryKineticsProteinLigand2015,
  title = {Kinetics of Protein\textendash Ligand Unbinding: {{Predicting}} Pathways, Rates, and Rate-Limiting Steps},
  shorttitle = {Kinetics of Protein\textendash Ligand Unbinding},
  author = {Tiwary, Pratyush and Limongelli, Vittorio and Salvalaglio, Matteo and Parrinello, Michele},
  year = {2015},
  month = feb,
  journal = {Proceedings of the National Academy of Sciences},
  volume = {112},
  number = {5},
  pages = {E386-E391},
  publisher = {{Proceedings of the National Academy of Sciences}},
  doi = {10.1073/pnas.1424461112}
}

@article{torrieNonphysicalSamplingDistributions1977,
  title = {Nonphysical Sampling Distributions in {{Monte Carlo}} Free-Energy Estimation: {{Umbrella}} Sampling},
  shorttitle = {Nonphysical Sampling Distributions in {{Monte Carlo}} Free-Energy Estimation},
  author = {Torrie, G.M. and Valleau, J.P.},
  year = {1977},
  month = feb,
  journal = {Journal of Computational Physics},
  volume = {23},
  number = {2},
  pages = {187--199},
  doi = {10.1016/0021-9991(77)90121-8}
}

@article{tribelloPLUMEDNewFeathers2014,
  title = {{{PLUMED}} 2: {{New}} Feathers for an Old Bird},
  shorttitle = {{{PLUMED}} 2},
  author = {Tribello, Gareth A. and Bonomi, Massimiliano and Branduardi, Davide and Camilloni, Carlo and Bussi, Giovanni},
  year = {2014},
  month = feb,
  journal = {Computer Physics Communications},
  volume = {185},
  number = {2},
  pages = {604--613},
  doi = {10.1016/j.cpc.2013.09.018}
}

@article{trizioEnhancedSamplingReaction2021,
  title = {From {{Enhanced Sampling}} to {{Reaction Profiles}}},
  author = {Trizio, Enrico and Parrinello, Michele},
  year = {2021},
  month = sep,
  journal = {The Journal of Physical Chemistry Letters},
  volume = {12},
  number = {35},
  pages = {8621--8626},
  doi = {10.1021/acs.jpclett.1c02317}
}

@article{valssonEnhancingImportantFluctuations2016,
  title = {Enhancing {{Important Fluctuations}}: {{Rare Events}} and {{Metadynamics}} from a {{Conceptual Viewpoint}}},
  shorttitle = {Enhancing {{Important Fluctuations}}},
  author = {Valsson, Omar and Tiwary, Pratyush and Parrinello, Michele},
  year = {2016},
  month = may,
  journal = {Annual Review of Physical Chemistry},
  volume = {67},
  number = {1},
  pages = {159--184},
  doi = {10.1146/annurev-physchem-040215-112229}
}

@article{valssonVariationalApproachEnhanced2014b,
  title = {Variational {{Approach}} to {{Enhanced Sampling}} and {{Free Energy Calculations}}},
  author = {Valsson, Omar and Parrinello, Michele},
  year = {2014},
  month = aug,
  journal = {Physical Review Letters},
  volume = {113},
  number = {9},
  pages = {090601},
  publisher = {{American Physical Society}},
  doi = {10.1103/PhysRevLett.113.090601}
}

@article{vargasConformationalStudyAlanine2002a,
  title = {Conformational {{Study}} of the {{Alanine Dipeptide}} at the {{MP2}} and {{DFT Levels}}},
  author = {Vargas, Rubicelia and Garza, Jorge and Hay, Benjamin P. and Dixon, David A.},
  year = {2002},
  month = apr,
  journal = {The Journal of Physical Chemistry A},
  volume = {106},
  number = {13},
  pages = {3213--3218},
  doi = {10.1021/jp013952f}
}

@article{voterHyperdynamicsAcceleratedMolecular1997,
  title = {Hyperdynamics: {{Accelerated Molecular Dynamics}} of {{Infrequent Events}}},
  shorttitle = {Hyperdynamics},
  author = {Voter, Arthur F.},
  year = {1997},
  month = may,
  journal = {Physical Review Letters},
  volume = {78},
  number = {20},
  pages = {3908--3911},
  doi = {10.1103/PhysRevLett.78.3908}
}

@article{wehmeyerTimelaggedAutoencodersDeep2018,
  title = {Time-Lagged Autoencoders: {{Deep}} Learning of Slow Collective Variables for Molecular Kinetics},
  shorttitle = {Time-Lagged Autoencoders},
  author = {Wehmeyer, Christoph and No{\'e}, Frank},
  year = {2018},
  month = mar,
  journal = {The Journal of Chemical Physics},
  volume = {148},
  number = {24},
  pages = {241703},
  doi = {10.1063/1.5011399}
}

\appendix
\setcounter{figure}{0}
\renewcommand{\thefigure}{S\arabic{figure}}
\section*{Supporting Information}
\FloatBarrier

\begin{figure}[h!]
\centering
\includegraphics[width=\linewidth]{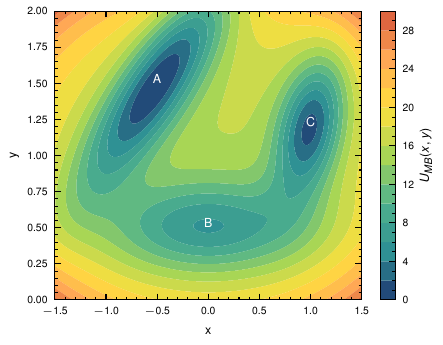}
\caption{Isolines of the 2D model potential energy surface.}
\end{figure}

\begin{figure}[h!]
\centering
\includegraphics[width=0.8\linewidth]{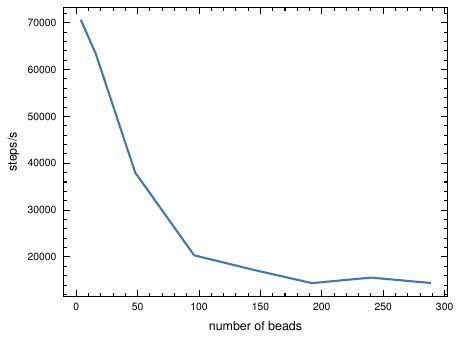}
\caption{Performance of MoP as a function of the number of beads in the polymer, measured in MoP steps per second, running with one bead per MPI task and one CPU core per bead. The performance decreases from ~70k steps/s for a very small polymer made of 4 beads to ~14k steps/s for a polymer made of 200 beads, after which a plateau is reached that extends up to the number of N=288 beads used in our simulations. Note that, while the performance decreases by a factor of 70/14=5, the number of beads in the polymer has grown by a factor of 288/4=72, which is more than one order of magnitude larger.}
\end{figure}

\begin{figure}
\centering
\includegraphics[width=\linewidth]{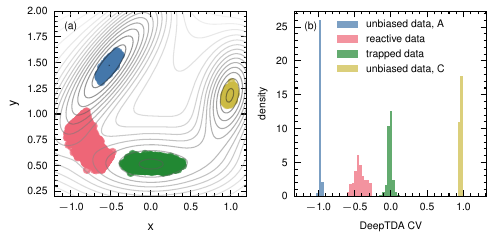}
\caption{Training data and histogram of CV distributions for training the 4-state DeepTDA CV in the case of the 2D model potential. (a) Scatter plot of the training data, with data from unbiased simulations in the left and right basins shown respectively in blue and yellow, selected configurations from reactive paths generated by MoP shown in pink and configurations from trapped paths in the intermediate basin shown in green. From all 4 states, 6666 configurations are shown and used to train the 4-state DeepTDA CV. Isolines of the 2D model potential are indicated in grey. (b) Histogram of the trained 4-state DeepTDA CV evaluated on the training data from the different basins. Note how the unbiased data are mapped to sharp peaks at the end points, and data from reactive and confined trajectories are mapped to broader distributions.}
\end{figure}

\begin{figure}
\centering
\includegraphics[width=\linewidth]{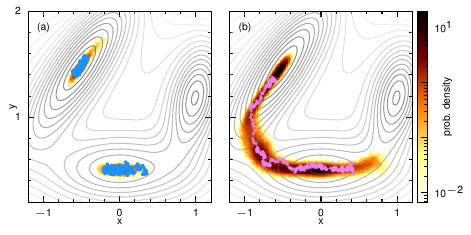}
\caption{Trapped (a) and reactive (b) trajectories, isolated from a MoP simulation using the 2-state DeepTDA CV for the 2D model potential in trajectory space. Examples of trapped trajectories are shown in blue, a reactive trajectory in pink. The densities are calculated from 152 trapped and 1855 reactive trajectories, respectively. Isolines of the potential are indicated in grey.}
\end{figure}

\begin{figure}
\centering
\includegraphics[width=\linewidth]{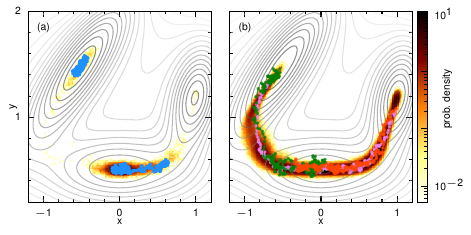}
\caption{Trapped (a) and reactive (b) trajectories, isolated from a MoP simulation using the 4-state DeepTDA CV for the 2D model potential in trajectory space. The densities are calculated from 208 trapped and 2603 reactive trajectories, respectively. Examples of trapped trajectories are shown in blue. Reactive trajectories connecting the basins A-B, A-C and B-C are shown in green, pink and red, respectively. Isolines of the potential are indicated in grey.}
\end{figure}

\begin{figure}
\centering
\includegraphics[width=\linewidth]{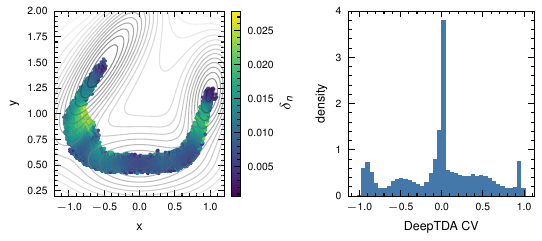}
\caption{Trajectories sampled by MoP spend more time in the vicinity of transition states. (a) Moving average of the oriented displacement between adjacent beads in a trajectory (average adjacent displacement, or AAD). If $\{\mathbf{d}_n\}_{n=1}^N$ is a set of descriptors or CV$_c$s, at bead $n$ this measure is given by $\delta_n = \left| \frac1{2k} \sum_{m=-k}^k \mathbf{d}_{n-m+1} - \mathbf{d}_{n-m} \right|$, where $k$ is the kernel width of the moving average. Thereby, it directly shows regions of \textit{increased} density along the polymer as regions of \textit{decreased} AAD. Here it is evaluated along the reactive trajectories obtained from a MoP simulation using the 4-state DeepTDA CV$_t$, as described in the main text, with $k=15$. Note that AAD is decreased not only in the known metastable states A and C, but also in state B and on the transition states connecting them. (b) Histogram of the 4-state DeepTDA CV, evaluated on the reactive paths obtained in a MoP simulation. Note the peaks in the distribution around $s=-1,1$ corresponding to basins A and C, the sharp peak at $s=0$ corresponding to the intermediate basin B, and peaks around $s=-0.5,0.5$, corresponding to the transition states.}
\end{figure}

\begin{figure}
\centering
\includegraphics[width=\linewidth]{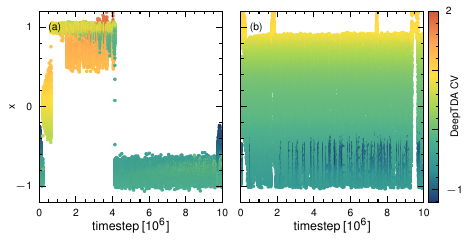}
\caption{Time evolution of the x coordinate of the 2D model potential, during biased simulations in configurational space using the (a) 2-state and (b) 4-state DeepTDA CVs, colored according to respective CV value. }
\end{figure}

\begin{figure}
\centering
\includegraphics[width=\linewidth]{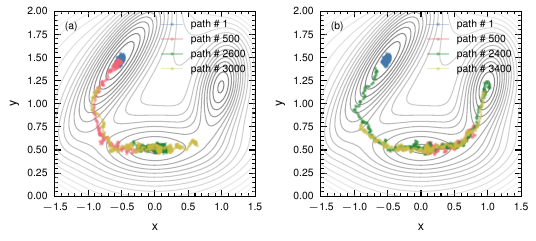}
\caption{Polymer snapshots of the MoP simulations using (a) the 2-state DeepTDA CV and (b) the 4-state DeepTDA CV on the 2D model potential. The equilibrated initial trajectory is shown in blue.}
\end{figure}

\begin{figure}
\centering
\includegraphics[width=\linewidth]{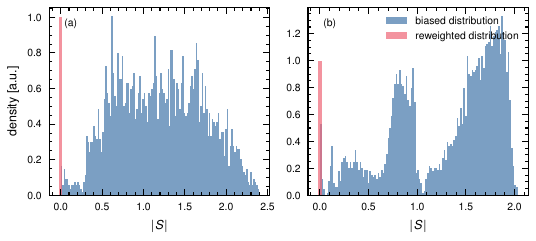}
\caption{The biased distribution of $|S|$ values in all polymers sampled by MoP using (a) the 2-state and (b) the 4-state DeepTDA CV on the 2D model potential is shown in blue. In panel (b), the structure centered at $|S|\approx 0.8$ corresponds to partial reactive paths connecting basin A to B or B to C of the 2D model potential potential, while the structure centered at $|S|\approx 1.9$ corresponds to complete reactive paths connecting A to C. In red, we report the distribution obtained after reweighting following the scheme of Ref.~\cite{invernizziRethinkingMetadynamicsBias2020a}, showing the expected delta-like peak around $|S|\approx 0$. The reweighted distribution has been scaled for clarity of presentation. We note that in this work, MoP was used in an exploratory manner with the sole purpose of generating reactive trajectories. Therefore, these distributions are not converged. 
}
\end{figure}

\begin{figure}
\centering
\includegraphics[width=\linewidth]{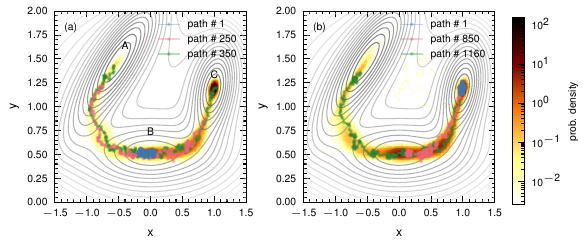}
\caption{MoP simulations on the 2D model potential starting from (a) basin B and (b) basin C, shown as normalized densities of all configurations in the polymers in logarithmic scale. Snapshots of selected trajectories are also shown, including the initial equilibrated trajectory (blue), and a complete reactive trajectory (green).}
\end{figure}

\begin{figure}
\centering
\includegraphics[width=\linewidth]{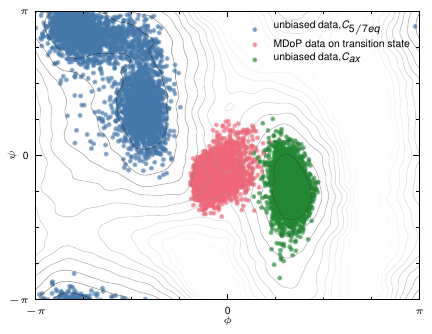}
\caption{Training data for the 3-state DeepTDA CV in the case of alanine dipeptide. Scatter plot of the training data, with data from unbiased simulations in the C$_{5,7eq}$ and C$_{ax}$ basins shown respectively in blue and green, and selected configurations from reactive paths generated by MoP shown in pink. From all 3 states, 4000 configurations are shown and used to train the 3-state DeepTDA CV. Isolines of the FES calculated from a converged reference calculation using the $\phi,\psi$ dihedral angles are indicated in grey.}
\end{figure}

\begin{figure}
\centering
\includegraphics[width=\linewidth]{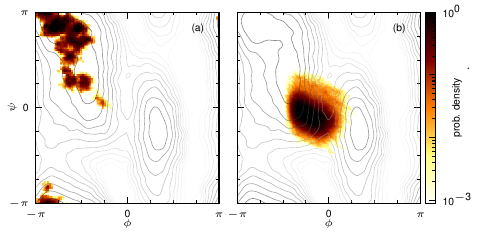}
\caption{Trapped (a) and reactive (b) trajectories, isolated from a MoP simulation using the 2-state DeepTDA CV for alanine dipeptide in trajectory space. The normalized densities are calculated from 125 trapped and 3529 reactive trajectories, respectively. Isolines of the FES of alanine dipeptide calculated from a converged reference calculation using the $\phi,\psi$ dihedral angles are indicated in grey.}
\end{figure}

\begin{figure}
\centering
\includegraphics[width=\linewidth]{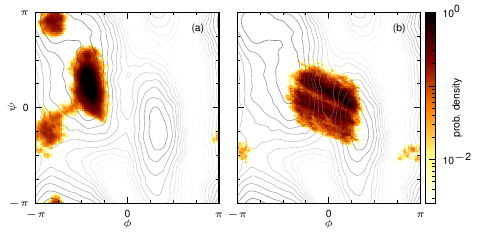}
\caption{Trapped (a) and reactive (b) trajectories, isolated from a MoP simulation using the 3-state DeepTDA CV for alanine dipeptide in trajectory space. The normalized densities are calculated from 1405 trapped and 569 reactive trajectories, respectively. Isolines of the FES of alanine dipeptide calculated from a converged reference calculation using the $\phi,\psi$ dihedral angles are indicated in grey.}
\end{figure}

\begin{figure}
\centering
\includegraphics[width=\linewidth]{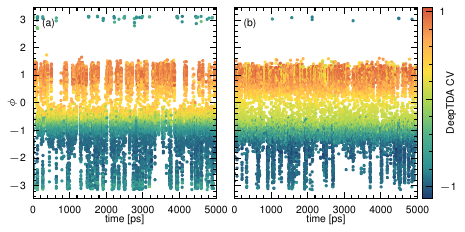}
\caption{Time evolution during biased simulations in configurational space using the (a) 2-state and (b) 3-state DeepTDA CVs, colored according to respective CV value.}
\end{figure}

\begin{figure}
\centering
\includegraphics[width=\linewidth]{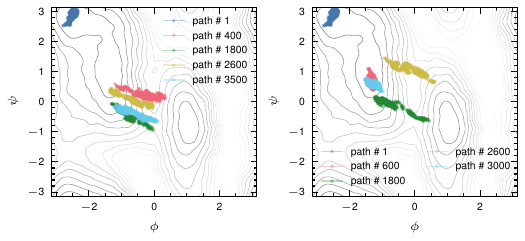}
\caption{Polymer snapshots of the MoP simulations using (a) the 2-state DeepTDA CV and (b) the 3-state DeepTDA CV on alanine dipeptide. The equilibrated initial trajectory is shown in blue.}
\end{figure}

\begin{figure}
\centering
\includegraphics[width=\linewidth]{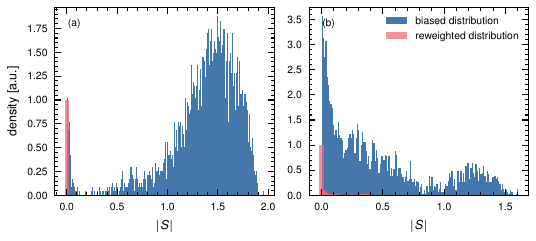}
\caption{The biased distribution of $|S|$ values in all polymers sampled by MoP using (a) the 2-state and (b) the 3-state DeepTDA CV on alanine dipeptide is shown in blue. In red, we report the distribution obtained after reweighting following the scheme of Ref.~\cite{invernizziRethinkingMetadynamicsBias2020a}, showing the expected delta-like peak around $|S|\approx 0$. The reweighted distribution has been scaled for clarity of presentation. We note that in this work, MoP was used in an exploratory manner with the sole purpose of generating reactive trajectories. Therefore, these distributions are not converged.}
\end{figure}

\begin{figure}
\centering
\includegraphics[width=\linewidth]{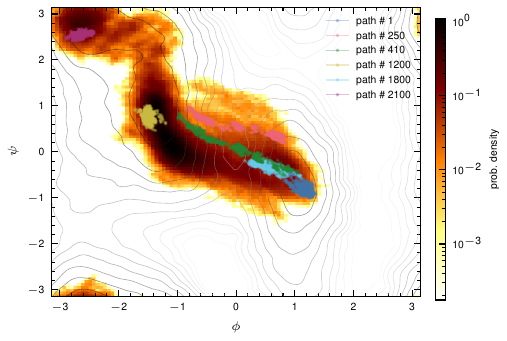}
\caption{MoP simulations on Alanine Dipeptide, starting from the Cax basin, shown as normalized densities of all configurations in the polymers in logarithmic scale. Snapshots of selected trajectories are also shown, including the initial equilibrated trajectory (blue), and a complete reactive trajectory (green).}
\end{figure}

\end{document}